\documentstyle[twoside,fleqn,espcrc2]{article}
\title{Bose Condensation and Superfluidity in Finite Rotating Bose Systems}
\author{Juhao Wu and A. Widom  
\address{Physics Department, Northeastern University, Boston MA 02115}
}
\begin{document}

\begin{abstract}
There is a long standing problem about how close a connection exists 
between superfluidity and Bose condensation. Employing recent 
technology, for the case of confined finite Bose condensed systems 
in TOP traps, these questions concerning superfluidity and Bose 
condensation can be partially resolved if the velocity profile of the 
trapped atoms can be directly measured.
\end{abstract}

\maketitle

\section{Introduction}

A long standing problem in the theory of superfluidity concerns the 
question of how close is the connection between Bose condensation 
and superfluidity\cite{1}. For the case of superfluid 
$^4He$, the Bose condensate fraction is quite small (if not virtually zero), 
yet the superfluid fraction goes to unity in the limit of low 
temperatures $T\to 0$. For the superfluid $^4He$ case, the observation 
of superfluidity is much more simple than the observation of Bose 
condensation. For the case of dilute Bose gases in atomic traps, it has 
been the case that the observation of Bose condensation 
appears more simple than the observation of superfluidity. 

Let us pause to review the differences in theoretical definition 
between the Bose condensate fraction and superfluid fraction. For a 
Bose fluid in equilibrium, the reduced one particle density matrix 
is defined as 
\begin{equation}
({\bf r}|\rho |{\bf r}')
=\big<\hat{\psi }^\dagger ({\bf r}')\hat{\psi }({\bf r})\big>.
\end{equation}
The one particle density matrix has a 
{\it maximum eigenvalue} $N_{max}$ defined by  
\begin{equation}
\int ({\bf r}|\rho |{\bf r}')\Psi ({\bf r}')d^3r'
=N_{max}\Psi ({\bf r}),
\end{equation} 
where $\Psi ({\bf r})$ is the Bose condensate wave function. 
The condensate fraction may then be defined as 
\begin{equation}
\eta_{c}=(N_{max}/N)
\end{equation}
where $N$ is the total number of atoms. 

Now suppose that the fluid were in a rotating bucket\cite{2}. Let 
${\bf \Omega }$ denote the angular velocity of the bucket.
If the fluid were to rotate as a rigid body, (as would any 
classical fluid), then the moment of inertia would be given by 
\begin{equation}
I_{ij}=\int <\hat{\rho}({\bf r})>\big(r^2\delta_{ij}-r_ir_j\big)d^3r,
\end{equation} 
where $<\hat{\rho }({\bf r})>$ is the mean mass density of the fluid.
If the fluid were to rotate as a superfluid, then the orbital angular 
momentum $<\hat{\bf L}>$ of the atoms would be somewhat smaller than 
that expected on the basis of rigid body rotation. Only the normal 
fluid rotates along with the bucket. This leads to the notion of a 
normal fluid fraction 
\begin{equation}
\eta_n=\Big(
{
{\bf \Omega \cdot }{<\hat{\bf L}>}
\over {\bf  \Omega \cdot I \cdot \Omega}
}
\Big).
\end{equation} 
The superfluid fraction in the Landau two fluid picture is defined as  
\begin{equation}
\eta_s =1-\eta_n
\end{equation}

Our purpose is to point out that {\it both} the condensate fraction 
$\eta_c $ {\it and} the superfluid fraction $\eta_s $ play a central 
role for Bose systems in TOP traps. The TOP trap homogeneous rotating 
contribution to the magnetic field serves to induce a laboratory 
rotating bucket. Thus, the question of whether or not a Bose 
condensate also produces superfluidity reduces to the question 
of whether or not the trapped fluid flow follows the rotation of the 
bucket. 

If the fluid does not circulate in the TOP trap, then the fluid 
exhibits superfluid behavior. If the fluid does rotate  
with a rigid body flow, closely following the bucket, then the fluid 
exhibits normal fluid behavior. If there is partial rotation of the 
fluid, then there is a finite superfluid fraction. 

\section{TOP Trap Magnetic Fields}

The Hamiltonian for atoms in a TOP trap is thought to have the time 
varying form  
\begin{equation}
H_{TOP}(t)=\sum_{1\le j\le N} h^{TOP}_j(t)+\sum_{1\le j < k\le N}v_{jk},
\end{equation}
where $v_{jk}$ is a two body potential, 
\begin{equation}
h^{TOP}_j(t)=-\Big({\hbar^2\over 2M}\Big)\nabla_j^2 -
\gamma_j {\bf S}_j {\bf \cdot B}({\bf r}_j,t) 
\end{equation}
is a one body Hamiltonian, $\gamma_j$ is the gyro-magnetic ratio 
of the $j^{th}$ atom, and ${\bf S}_j$ is the total spin of the 
$j^{th}$ atom.

The time varying magnetic field in a TOP trap obeys  
\begin{equation}
{\bf B}({\bf r},t)={\bf B}_h(t)+{\bf B}_Q({\bf r}),
\end{equation}
where the quadrapole field contribution ${\bf B}_Q({\bf r})$ is given 
in terms of a static field gradient amplitude $G$, 
\begin{equation}
{\bf B}_Q({\bf r})=G\big({\bf r}-3({\bf n\cdot r}){\bf n}\big).
\end{equation}
The homogeneous and time varying field ${\bf B}_h(t)$ rotates about 
(and is normal to) the unit vector ${\bf n}$; i.e. the 
``bucket angular velocity'' is of the form 
${\bf \Omega }=\Omega {\bf n}$. Furthermore, 
${\bf n\cdot B}_0=0$ and  
\begin{equation}
{\bf B}_h(t)={\bf B}_0 \cos(\Omega t)+
{\bf n\times B}_0 \sin(\Omega t). 
\end{equation} 

To understand why TOP trap magnetic field induces the Hamiltonian of 
a ``rotating bucket'', it is sufficient to introduce the total angular 
momentum, (i.e. total orbital plus total spin) of all $N$ atoms 
\begin{equation}
\hat{\bf J}=\hat{\bf L}+\hat{\bf S}
=\sum_j \big({\bf r}_j{\bf \times p}_j\big)+\sum_j{\bf S}_j ,
\end{equation}
where ${\bf p}_j=-i\hbar {\bf \nabla}_j$. We employ the unitary 
rotation operator  
\begin{equation}
U(t)=\exp\Big(-{i{\bf \Omega \cdot }{\hat{\bf J}t}\over \hbar }\Big)
=\exp\Big(-{i\Omega {\bf n \cdot }{\hat{\bf J}t}\over \hbar }\Big),
\end{equation}
within the canonical transformation $H_{TOP}(t)\to {\cal H}$, as 
given by
\begin{equation}
{\cal H}=U^\dagger (t)H_{TOP}(t)U(t)
-i\hbar U^\dagger (t)\Big({\partial U(t)\over \partial t}\Big).
\end{equation}
This transformation leads to a time independent Hamiltonian 
\begin{equation}
{\cal H}=\hat{H}-{\bf \Omega \cdot }\hat{\bf J}.
\end{equation}
In the above Eq.(15), the time independent contribution $\hat{H}$ 
is given by 
\begin{equation}
\hat{H}=\sum_{1\le j\le N} h_j+\sum_{1\le j < k\le N}v_{jk},
\end{equation}
where
\begin{equation} 
h_j=-\Big({\hbar^2\over 2M}\Big)\nabla_j^2 -
\gamma_j {\bf S}_j {\bf \cdot }\tilde{\bf B}({\bf r}_j). 
\end{equation}
In the rotating frame, the magnetic field operator may be 
viewed as being static; i.e.  
\begin{equation}
\tilde{\bf B}({\bf r})={\bf B}_0+
G\big({\bf r}-3({\bf n\cdot r}){\bf n}\big).
\end{equation}
Thus Eqs.(15), (16), and (17) determine the typical ``rotating  
bucket'' form for the TOP trap Hamiltonian. 

\section{The Rotating Bucket Hamiltonian}

As previously stated, The Hamiltonian in Eq.(15) is precisely of that 
type which is found for Bose fluids in rotating buckets. However, 
TOP traps have not always been viewed in this fashion. In the original 
experimental work on Bose condensates in TOP traps\cite{3,4}, one employed 
a time averaged adiabatic Hamiltonian in the analysis. 
Such an analysis may hold true if and only if 
the bucket rotates and the contained fluid does not rotate; 
i.e. such a model may hold true but only if the fluid acted as a perfect 
superfluid. The superfluid fraction in Eqs.(5) and (6) would have to obey  
$\eta_s=1$ in order for the time averaged adiabatic Hamiltonian view to 
have any merit. In this time averaged view, the bucket rotates and the 
fluid does not rotate, a very perfect superfluid indeed.  

It is very unlikely that such a perfect superfluid condition 
works for realistic experiments with the substantially high rotational 
velocities of TOP traps. There is, in fact, no compelling physical 
motivation to ever time average the Hamiltonian. 
Time averaging is not the proper way to treat fluids in rotating 
buckets. 

In more recent work\cite{5,6,7,8}, the time averaging was still being 
carried out but the rotational velocity was properly included only in 
so far as the spin degrees of freedom are concerned. The effective 
Hamiltonian of this most recent work reads\cite{9} (in our notation) 
as
$$
{\cal H}_{eff}=\hat{H}-{\bf \Omega \cdot }\hat {\bf S}. \eqno(\rm Wrong)
$$
The above equations is labeled as being wrong only by reason of employing 
a partial transformation (rotating only in spin angular momentum and 
not in orbital angular momentum). Such a transformation leaves an 
effective Hamiltonian which still depends on time by virtue of the 
quadrapole contribution to the magnetic field. Only after time averaging 
can the explicit time dependence be removed from ${\cal H}_{eff}$. The 
weakness of such an approach has been noted.

The proper method employs Eq.(15), in which the total angular momentum 
(spin plus orbital) generates the rotations.  The resulting rotating frame 
Hamiltonian ${\cal H}$ is truly time independent, and properly includes 
the coupling to the orbital angular momentum of the atoms. 
This orbital angular momentum has been previously discussed
\cite{10} for TOP traps. The above Eq.(Wrong) should be replaced by
$$
{\cal H}=\hat{H}-{\bf \Omega \cdot }\big(\hat{\bf S}+\hat{\bf L}\big). 
\eqno(\rm Correct)
$$
The above coupling is crucial for a proper analysis of rotational 
superfluidity. This is quite important for determining the nature of 
the connection between the superfluid fraction of Eqs.(5) and (6) 
and the Bose condensate fraction of Eq.(3). 

Finally, in the adiabatic approximation, Eqs.(15), (16), (17) and (18) 
imply Hamiltonians of the form 
\begin{equation}
{\cal H}_{adiabatic}=\hat{H}_{ad}-
{\bf \Omega \cdot }\hat{\bf L}
\end{equation}
where 
\begin{equation}
\hat{H}_{ad}=\sum_{1\le j\le N} h^{ad}_j+
\sum_{1\le j < k\le N}v_{jk},
\end{equation}
and where a simple adiabatic potential model yields    
\begin{equation}
h^{ad}_j=-\Big({\hbar^2\over 2M}\Big)\nabla_j^2 
+V^{ad}_j({\bf r}_j). 
\end{equation}
The adiabatic potential is given by  
\begin{equation}
V^{ad}_j({\bf r}_j)=
\big|M_{S,j}\gamma_j\big|
\Big|\tilde {\bf B}({\bf r}_j)+{{\bf \Omega }\over \gamma_j}\Big|,
\end{equation}
where $\tilde{\bf B}({\bf r})$ is defined in Eq.(18) and $M_{S,j}$ is an 
appropriate projection of the $j^{th}$ atom total spin. Had we performed 
the adiabatic approximation before the rotational unitary transformation  
(as in previous work\cite{10}), the adiabatic time independent 
rotational Hamiltonian would {\it still} have the form of Eq.(19). 
However, the adiabatic potential in Eq.(22) would be missing the spin 
induced ${\bf \Omega}/\gamma_j$ term. The subtle point is that 
the adiabatic limit and the rotational unitary transformation are not 
quite commuting processes. This leads to corrections in the adiabatic 
results which start at order $|\Omega /(\gamma_j B_0)|$. 

\section{Magnitudes} 

Let us consider the two rotational terms in 
\begin{equation}
{\cal H}=\hat{H}-{\bf \Omega \cdot }\hat{\bf J}=
\hat{H}-{\bf \Omega \cdot }\big(\hat{\bf S}+\hat{\bf L}\big).
\end{equation}
The spin term is at most 
$|{\bf \Omega \cdot }<\hat{\bf S}>|\sim N\hbar \Omega $. Furthermore, if 
there were but one single vortex line in the Bose condensed system, then 
the orbital term would also obey 
$$
|{\bf \Omega \cdot }<\hat{\bf L}>|\sim N\hbar \Omega . 
\eqno(\rm One\ Vortex\ Line)
$$
With one (or less) vortex line, the trapped Boson system would constitute  
an almost perfect superfluid in the sense of Eqs.(5) and (6). 

On the other hand, the angular velocity of the TOP trap rotating bucket  
is high on the scale of simple estimates of the critical velocity, i.e. 
\begin{equation}
\Omega >>\Omega_c\sim \Big({\hbar \over M b^2}\Big)
\end{equation} 
where $b$ is a sensible length scale describing the size of the bucket. 
Employing such a simple estimate, one would expect many vortices. 
The formation of many vortices implies 
$$
|{\bf \Omega \cdot }<\hat{\bf L}>|>>|{\bf \Omega \cdot}<\hat{\bf S}>|. 
\eqno(\rm Many\ Vortices)
$$
Under no circumstances (other than vortex free perfect superfluidity)  
may the orbital angular momentum rotational 
coupling be regarded as a small perturbation on the spin 
rotational angular momentum coupling.  This situation has been incorrectly 
described elsewhere\cite{6}.

Only for the case of a unit superfluid fraction $\eta_s\approx 1$ 
will there be zero vorticity at large bucket rotational velocities. 
In previous work\cite{11}, Rokhsar stated that from the work of Putterman, 
Kac, and Uhlenbeck\cite{12}, it follows that vortices in TOP traps will have 
cores which are not pinned. The vortices can thereby slip out of the fluid. 
Such depinning would leave behind a non-rotating condensate. 

A non-rotating condensate together with a quickly rotating bucket means 
{\em perfect superfluidity}. For the continually rotating bucket 
situation with $\Omega >>\Omega_c$ there may be {\it  stable vortices}. 
In judging the vortex stability one should use a statistical ensemble 
in which ${\bf \Omega }$ is fixed and angular momentum fluctuates. This 
is especially true when angular momentum is not conserved by virtue 
of the bucket potential which is not rotationally 
invariant\cite{12}. It is unreliable for estimates of stability to employ  
the fixed angular momentum ensemble as in the work of Rokhsar\cite{13}. 
The idea of an absolutely rotationally invariant bucket is a mathematical 
fiction not physically present in laboratories. It has long been known 
since the time of Newton that if you rotate a bucket fast enough, then 
the fluid will also rotate. 

The angular velocity of the bucket can induce the vortices into a stable  
(rather than a metastable) configuration. The higher the TOP trap angular 
velocity, the more stable are the vortices. In this manner, 
rotating buckets containing a superfluid may yield a 
simulation (i.e. imitation) of a normal fluid. This requires stable 
vortex formation. In a fully Bose condensed system with $\eta_c\approx 1$, 
and with a mass current determined by the condensate wave function 
$\Psi ({\bf r })$, the formation of many vortices are the only means 
available for simulating the rigid body rotation within the fluid.
Such an effects are well known for superfluid $^4He$ case even 
though $\eta_c(^4He)<<1$.

\section{Conclusions}

A fluid which undergoes rigid body rotation has a velocity field of 
the form 
\begin{equation}
{\bf v}({\bf r })={\bf \Omega }\times {\bf r}.
\end{equation}
A classical fluid in a rotating bucket always\cite{2} exhibits a 
velocity field as in Eq.(25). With the phase $\phi $ of the superfluid 
determined by the condensate wave function 
$\Psi =|\Psi |e^{iM\phi/\hbar}$ , the superfluid velocity 
\begin{equation}
{\bf v}_s ({\bf r})={\bf \nabla}\phi ({\bf r})
\end{equation} 
is irrotational and cannot duplicate the rotational velocity field 
of Eq.(25). The best that can be done is to create many vortices at 
a density which (on a coarse grained average) can simulate rigid body 
rotation and thus appear to be ``normal''.

The final analysis is experimental and not theoretical. If, upon 
measuring the velocity profile of a TOP trap Bose system (via the 
Doppler shift) one observes a velocity field of the form in Eq.(25), 
then the fluid is ``normal'' for the high ${\bf \Omega }$, albeit 
in a possible Bose condensed state. If zero velocity field is detected 
in a rotating TOP trap, then the condensate is superfluid, and is 
in fact quite remarkable. It is unusual to observe a significant 
superfluid fraction in the regime $\Omega >>\Omega_c $ of Eq.(24).

\end{document}